\documentclass[12pt]{article}
\usepackage{bm}
\usepackage{amsmath}
\usepackage{amssymb}
\usepackage{graphicx}
\usepackage{amsmath}
\usepackage{bbm}

\begin{document}

\baselineskip 15pt

\title{Quantum Mechanics and Perceptive Processes: a reply to Elio Conte}

\author{GianCarlo Ghirardi\footnote{e-mail: ghirardi@ts.infn.it}\\ {\small
Emeritus, Department of  Physics, the University of Trieste,}\\ {\small the Abdus
Salam International Centre for Theoretical Physics, Trieste, Italy.}}

\date{}

\maketitle

\begin{abstract}
Recently, Elio Conte has commented a paper by the present author devoted to analyze the possibility of checking experimentally  whether the perceptual process can lead to the collapse of the wavefunction. Here we answer to the comments by Conte and we show that he has missed to grasp the crucial elements of our proposal. Morever, we discuss some ideas put forward by Conte concerning the occurrence of quantum superpositions of different states of consciousness and we show that they are rather vague and not  cogent. 
\end{abstract}

\vspace{2cm}

\section{Introduction} 

We have read with interest the paper \cite{conte} by Elio Conte: {\it Answer to GianCarlo Ghirardi}  in which he discusses some  experiments which are relevant for  Perceptive and Cognitive Sciences. Here, we will reply to his remarks on our paper \cite{ghirardi}   and we will analyze critically various other arguments presented by him.

The most relevant fact, for what concerns our contribution, is that  in ref.\cite{conte} our position has been to a large extent misunderstood. In fact, the question we have tackled is whether the perceptual apparatus is able to induce {\bf the suppression of linear superpositions} of states each of which is associated to a different perception and whether this fact might be experimentally investigated. The aim is obvious, i.e. to account for our definite perception even in the case in which the stimulus is generated by a microsystem in a superposition of two states which, by themselves, would induce different perceptions. 

On the contrary, ref.\cite{conte} addresses exclusively the quite interesting problem of {\bf whether quantum superpositions of brain states associated to different perceptions occur} and its answer to this fundamental question seems to be a clear-cut YES. In a sense we are interested in opposite problems: we believe that the perceptive processes should induce reduction of quantum superpositions and we have suggested a way to check this experimentally, while the author is concentrated on the {\bf actual} occurrence of states corresponding to different perceptions and to experimental investigations on this matter.

The arguments which are used by Conte for deriving his conclusions concerning  superpositions of different perceptions seem to us rather vague and inconsistent.  Accordingly, we will devote part of this paper to discuss them in detail.

\section{Conte's position concerning our paper} According to  ref.\cite{conte} the investigations on Quantum Cognition have put into evidence that quantum mechanics plays a basic role during the perceptive and cognitive performances in humans. This  claim is based on the assertion that  the author and other scientists\footnote{For these researches we refer the reader to the rich bibliography of ref.\cite{conte}.} have reached decisive experimental evidence of the occurrence of quantum interference during our perceptive-cognitive performances.

Let me start by considering some arguments which should support such crucial experimental evidence. After having reported our argument in pages 2 and 3 of the paper, the author, to reply to our proposal which, I insist, aimed to understand  whether the perceptual process can lead to the collapse of the statevector, starts by stressing that {\it the binocular and biaural hearing rise interesting features and questions, ..., in particular at the level of consciousness}.

My first remark is that to make reference  to the binocular vision, as if it would play any role in our analysis, is fully inappropriate.  We have considered  perceptions corresponding to a luminous spot from a space point A or to a luminous spot from a space point B, A and B being at a macroscopic distance (many centimeters) from each other. Now, it goes without saying that the perceptions {\it the light comes from A} or {\it the light comes from B} are different and can very well be distinguished also by a person who has one of his eyes closed or in some way screened from the source: one  eye by itself is certainly sufficient to trigger different perceptions in the two cases. Accordingly, invoking binocular vision is totally irrelevant for the argument.

 Before proceeding, we consider it useful to add  a specific comment. In a debate like the present one we  definitely consider more appropriate to make reference, as we did in our paper, to the preceptive process than to  consciousness. In fact, the whole program of reduction by consciousness championed by J. von Neumann and subsequently by E. P. Wigner, has been  abandoned by Wigner himself as lucidly stressed by  M. Esfeld \cite{esfeld} in his essay on {\it Wigner's view of Physical Reality} in which one reads:
 
 \begin{quote}
 {\it An analogous consideration applies to WignerÕs later change of mind. His idea contains a
viable option: If one considers it to be inappropriate to take recourse to the mind or
consciousness of an observer in the interpretation of quantum mechanics and if one regards
state reductions as objective physical events, it is reasonable to envisage a modification of the
Schr{\" o}dinger dynamics. The aim then is to achieve a more general dynamics that encompasses
state reductions. The most elaborate suggestion in this respect goes back to Ghirardi, Rimini
and Weber \cite{ghirardi2}.}
\end{quote}

\section {Discussing the arguments and the data of ref.\cite{conte}}
The author proceeds by recalling that {\it ``we have recently produced some experiments evidencing as  brain entrainment using photo and audio stimulation is able to enhance the activity of our ... nervous system...'' }\footnote {A remark. The author should be more specific about the states triggering the mentioned processes and evidences by making clear, first of all, whether the stimulus is  a quantum superposition of states leading to different perceptions, or a state corresponding to a unique, definite perception. In fact, the assumption that the initial state is a superposition  represents the starting and basic point of our analysis and, in our opinion, is of fundamental importance for our and his analysis.}.

At this point the author concentrates his attention on the problem of whether we can see single photons, and he anticipates a positive answer to this question, even though, at the end, he is compelled to take into account a number of photons practically equal to the one used in our paper, i.e. 9-10, hitting the eye in a time interval smaller than 100ms. To deal with this question the author appropriately analyzes the different behaviours of rods and cones in the visual process and describes how they work differently in dependence of the environment (illuminated or dark) in which the perceiving subject is situated. This part as well as the description of scotopic vision is interesting and informative.

From my point of view the two facts that about 9 photons are actually required at the receptors and  that the rods must respond to single photons even when the subject is not able to see such photons because they arrive too infrequently, are quite illuminating. Actually, as already stated, the first remark  makes clear why I have chosen such a number of photons (which, and this is absolutely crucial, have to be prepared in a quantum superposition). For what concerns the second fact I 
stress that the property of rods to  respond to single photons  is commonly assumed by all quantum physicists. The most clear example can be found in a paper\cite{adler} by S. Adler who discusses, with reference to the collapse models recently introduced\cite{ghirardi2}, the changes taking place in a rod due to the absorption of a photon.

The author of ref.\cite{conte}, calls  attention also to another fact which undoubtedly has a remarkable interest for people interested in perceptions, but which is not relevant for the problem at hand. He reconsiders the binocular vision and points out that when a subject is in a dark room and the experimental setup is arranged in such a way that the signal reaches one and only one of his eyes, he is normally unaware concerning which eye has actually seen the signal.

At this point, and with specific reference to the just mentioned fact, the author presents what he considers the central result of his paper: {\it in this condition we are forced to acknowledge that there is a state of ``consciousness'' which corresponds to a coherent superposition of seing the signal by the right eye and seeing the same signal by the left eye}.

In my opinion the statement ``we are forced ...'' is logically unfounded. But this is not the whole story. The just depicted situation is absolutely peculiar: the stimulus should induce a quantum superposition of  macroscopically different states of the perceptual apparatus. But we have unique and unambiguous perceptions.

The previous analysis  should have made clear that the statements of the author that there is no need to perform sophisticated experiments as the one proposed by us does not follow in any way whatsoever. Actually,  as we are going to discuss in the next section, the arguments of the author are vague and logically obscure, and, as pointed out from the very beginning of this paper, they have nothing to do with  what we have  discussed  within a rigorous quantum framework.

\section{The last part of the paper}

Our critical remarks are strengthened by the conclusive part of ref.\cite{conte}. Here, the author takes into account the case of the celebrated ambiguous figures which can be perceived in different ways. The best known example is the one of the ``profiles and cup'' proposed around 1915 by E. Rubin \cite{rubin}, which we present in Fig.1. 

\begin{figure}[t]
\centering
\begin{center} 
 \includegraphics{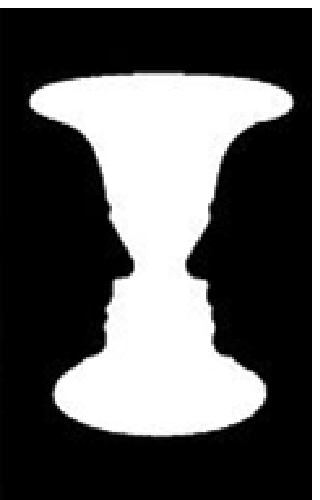} \\
 \caption{\footnotesize The famous image by E. Rubin}
\end{center}
\end{figure}

It is perfectly true (and well known) that one cannot perceive simultaneously the two images and that under continuous exposition of the figure the interpretation often oscillates back and forth. However, there is no reason  to deduce from these facts, as the author does, that one has to accept that at some point the consciousness leads to a quantum superposition of the two modes. Once more the argument is very obscure. 

In the  situation under consideration we have a classical object, only one, which remains identical to itself forever with the black profiles and the white cup. In no sense whatsoever this situation can be claimed to correspond, actually, to the  superposition of a page with the profiles and one with the cup.  To claim that the unicity of the image perceived at any given time and the changing in time of our perception proves that superpositions (in the precise quantum sense) of conscious states actually occur is something  unjustified and arbitrary.

In spite of this, let us accept tentatively that actually the perceptive status of the subject is a quantum superposition and let us present now a rigorous critical analysis of the problem by following  the  arguments of the author  looking at them from a correct quantum perspective\footnote {We will disregard the specific case considered at the beginning by the author, in which in the first experiment there is equal probability of getting any of the two possible perceptions and in the second there are probabilities $\sin^{2}\alpha$ and $\sin^{2}\beta$ for the subsequent outcomes and we will pass immediately to  the general case discussed by the author in which such probabilities take general  values.}.

The author considers the case in which test subjects look at a screen on which two ambiguous figures, B and A, are projected, one after the other. The figures admit only two possible interpretations, amounting, within the quantum context, to accept that we are dealing with dichotomic quantum observables. B and A play the role of two quantum (noncommuting) observables. For simplicity we will indicate as $+1$ and $-1$ their eigenvalues and as $\{|b=+\rangle, |b=-\rangle\}$ and $\{|a=+\rangle, |a=-\rangle\}$  the corresponding eigenvectors. The author of ref.\cite{conte}, even though it is not fully explicit on this point, assumes that immediately after being exposed to figure B the subject ``is'' in the quantum state:

\begin{equation}
|\psi_{initial}>=[\sqrt{p(B=+1)}\vert b=+\rangle+e^{i\theta}\sqrt{p(B=-1)}\vert b=-\rangle],
\end{equation}

\noindent where we have denoted as $p(B=+1)$ and $p(B=-1)$ the probabilities that the subsequent perception of the ambiguous figure B corresponds to $+1$ or to $-1$, respectively. Any phase factor in front of the whole state is obviously irrelevant.

At this point one of the two indicated perceptions emerges, which, within a quantum framework, means that reduction takes place and leads either to the state $\vert b=+\rangle$ or to the state $\vert b=-\rangle$ with the indicated probabilities.

And now comes the crucial question to which the author does not give an explicit answer but, at the same time,  makes very clear with his specifications, what he has in mind. Actually he considers now the exposition of the subject to the second ambiguous figure, and makes the following (rephrased to be completely general) statement\footnote {We make reference for the moment to the case in which after the second experiment  the subject perceives $A=+1$.}: {\it Let us admit now that the probability of the subjects' choosing A=+1, if they are sure that the first ambiguous figure means $B=+1$ is $p(A=+1|B=+1)$ while the probability of their choosing A=+1 if they are sure that the first figure means B=-1, is $p(A=+1|B=-1)$.}

If we stay strictly within a quantum context the above sentence means that, if the state of the subject is the eigenstate $|b=+\rangle$ of B,  then, looking at A, there is the probability $p(A=+1|B=+1)$ of perceiving A=+1, while if he is in the eigenstate  $|b=-\rangle$, there is a probability $p(A=+1|B=-1)$ of having such a perception.

If one follows this line of thought within a quantum scenario one can easily evaluate the probability $p(A=+1)$ of getting the final perception A=+1. In fact, in the time interval between observing the first and the second image one has a definite perception concerning the first image, which obviously means that he ends up in the state $|b=+\rangle$ or $|b=-\rangle$  with the indicated probabilities. Thus, quantum-mechanically we end up with a statistical mixture of subjects in the two indicated eigenstates. Since, in turn, when one is in one of these states one has the conditional probabilities $p(A=+1|B=+1)$ and $p(A=+1|B=-1)$ of getting the outcome A=+1 (which, incidentally means to assume that $|\langle a=+|b=+\rangle|^{2}=p(A=+1|B=+1)$, and the analogous one for the other case), following the quantum rules when dealing with a statistical mixture one ends up with the fact that $p(A=+1)$ takes precisely the form that the author derives using the classical Bayes theorem. It goes without saying that the same argument leads to the Bayes result for $p(A=-1)$.

In contradiction with his previous statement {\it if they are sure that the first ambiguous figure means B=+1 ...}, which quantum mechanically can mean only {\it if they are in the eigenstate $|b=+1\rangle$}, he considers that the state of the subject is still the one of Eq.(1). Obviously it makes no sense, in such a state, to say that {\it they are sure of ... }. In spite of this fact, let us take for granted   that, after being exposed to the observation of the ambiguous figure B the subjects are still in the state of Eq.(1). Then, to evaluate the probability $p(A=+1)$ of getting, after the second ambiguous image has been shown to them the perception A=+1 is given by the modolus square of the scalar product of the states $\vert a=+\rangle$ and $|\psi_{initial}\rangle$. If one evaluates this quantity, one easily gets the result: 

\begin{eqnarray}
p(A=+1)& =&p(B=+1)p(A=+1|B=+1)+p(B=-1)p(A=+1|B=-1)+\nonumber \\
&2&\sqrt{p(A=+1)p(B=-1)p(A=+1|B=+1)p(A=+1|B=-1)}\cos \theta
\end{eqnarray}

\noindent which is precisely the equation that the author claims to follow from the assumption that quantum superpositions occur into the brain.

The competent reader will have already grasped the reasons for which I can claim that the arguments of ref.\cite{conte} are mistaken and not understandable. If one takes the statement {\it if they are sure that the first ambiguous figure means $B=+1$} and similar ones, as indicating that reduction has taken place and thus the state must be an eigenstate of B then the computation gives the classical result. On the contrary, if one assumes that after having observed the first ambiguous figure the state of the subject is still the one of Eq.(1), then, quantum mechanically, it makes no sense to claim that {\it the subject is sure} concerning his first perception. 

The conclusion of this section should be obvious:  the arguments of ref.\cite{conte} referring to ambiguous figures are contradictory. The experiments that the author cites repeatedly as new and extremely relevant might be actually so, but they do not contribute in any way whatsoever to clarify the problems of whether perceptual processes are classical or quantum and of whether these processes play a precise role in the reduction process.

\section{Conclusion}
Summarizing, we stress once more that the paper cannot be considered in any sense an answer to ref.\cite{ghirardi} just because it addresses a quite different question than the one raised by us.  Moreover, all arguments which, in the author's perspective, would ``prove'' that quantum superpositions of conscious states actually occur, are contradictory and not amenable to any known physical process and/or accepted theoretical framework.

\section*{Acnowledgment}

We thank Prof. Catalina Curceanu for calling to our attention the paper by Conte.


\begin{thebibliography}{99}

\bibitem{conte}  Conte E. :{\it Int J Theor Phys}, DOI 10. 1007/s10773-014-2259-6.

\bibitem{ghirardi} Ghirardi G.C.: {\it Phys. Lett.} A {\bf 262}, 1 (1999).

\bibitem{esfeld} Esfeld M., {\it Stud. His. Phil. Modern Phys.}, {\bf 30 B}, 145 (1999).

\bibitem{ghirardi2} Ghirardi G.C., Rimini A. and Weber T., {\it Phys. Rev.} D{\bf 34}, 470 (1986).

\bibitem{adler} Adler S., {\it J. Phys.} A{\bf 40}, 2935, 13501 (2007).

\bibitem{rubin} E.Rubin, {\it Synsoplevede Figurer}, 1915.

\end{thebibliography}
\end{document}